\documentclass[12pt,preprint]{aastex}
\def\lessim{\mathrel{\hbox{\rlap{\hbox{\lower4pt\hbox{$\sim$}}}\hbox{$<$}}}}
\def\grtsim{\mathrel{\hbox{\rlap{\hbox{\lower4pt\hbox{$\sim$}}}\hbox{$>$}}}}

\shorttitle{M33 Novae}
\shortauthors{Shafter et al.}

\begin{document}

%% LaTeX will automatically break titles if they run longer than
%% one line. However, you may use \\ to force a line break if
%% you desire.

\title{On the Spectroscopic Classes of Novae in M33}

%% Use \author, \affil, and the \and command to format
%% author and affiliation information.
%% Note that \email has replaced the old \authoremail command
%% from AASTeX v4.0. You can use \email to mark an email address
%% anywhere in the paper, not just in the front matter.
%% As in the title, use \\ to force line breaks.

\author{A. W. Shafter\altaffilmark{1}, M. J. Darnley\altaffilmark{2}, M. F. Bode\altaffilmark{2}, and R. Ciardullo\altaffilmark{3}}
\altaffiltext{1}{Department of Astronomy, San Diego State University, San Diego, CA 92182, USA}
\altaffiltext{2}{Astrophysics Research Institute, Liverpool John Moores University, Birkenhead CH41 1LD, UK}
\altaffiltext{3}{Department of Astronomy and Astrophysics, The Pennsylvania State University, 525 Davey Lab, University Park, PA 16802, USA}
%\email{aws@nova.sdsu.edu, mjd@astro.livjm.ac.uk, mfb@astro.livjm.ac.uk, rbc@astro.psu.edu}

%% Notice that each of these authors has alternate affiliations, which
%% are identified by the \altaffilmark after each name.  Specify alternate
%% affiliation information with \altaffiltext, with one command per each
%% affiliation.

%% Mark off your abstract in the ``abstract'' environment. In the manuscript
%% style, abstract will output a Received/Accepted line after the
%% title and affiliation information. No date will appear since the author
%% does not have this information. The dates will be filled in by the
%% editorial office after submission.

\begin{abstract}
We report the initial results from an ongoing multi-year spectroscopic
survey of novae in M33. The survey resulted in the spectroscopic
classification of six novae (M33N 2006-09a, 2007-09a, 2009-01a,
2010-10a, 2010-11a, and 2011-12a) and a determination of rates
of decline ($t_2$ times) for four of them (2006-09a, 2007-09a, 2009-01a,
and 2010-10a). When these data are combined with existing spectroscopic data
for two additional M33 novae (2003-09a and 2008-02a) we find that five
of the eight novae with available spectroscopic class appear to be members
of either the He/N or Fe~IIb (hybrid) classes, with only two clear
members of the Fe~II spectroscopic class. This initial finding
is very different from what would be expected based on the results for
M31 and the Galaxy where Fe~II novae dominate, and the He/N
and Fe~IIb classes together make up only $\sim 20$\% of the total.
It is plausible that the increased fraction of He/N and Fe~IIb novae
observed in M33 thus far may be the result of the younger stellar population
that dominates this galaxy, which is expected to produce novae
that harbor generally more massive white dwarfs than those typically
associated with novae in M31 or the Milky Way.

\end{abstract}

\keywords{galaxies: stellar content --- galaxies: individual (M33) --- stars: novae, cataclysmic variables}

%% From the front matter, we move on to the body of the paper.
%% In the first two sections, notice the use of the natbib \citep
%% and \citet commands to identify citations.  The citations are
%% tied to the reference list via symbolic KEYs. The KEY corresponds
%% to the KEY in the \bibitem in the reference list below. We have
%% chosen the first three characters of the first author's name plus
%% the last two numeral of the year of publication as our KEY for
%% each reference.

%% Authors who wish to have the most important objects in their paper
%% linked in the electronic edition to a data center may do so by tagging
%% their objects with \objectname{} or \object{}.  Each macro takes the
%% object name as its required argument. The optional, square-bracket 
%% argument should be used in cases where the data center identification
%% differs from what is to be printed in the paper.  The text appearing 
%% in curly braces is what will appear in print in the published paper. 
%% If the object name is recognized by the data centers, it will be linked
%% in the electronic edition to the object data available at the data centers  
%%
%% Note that for sources with brackets in their names, e.g. [WEG2004] 14h-090,
%% the brackets must be escaped with backslashes when used in the first
%% square-bracket argument, for instance, \object[\[WEG2004\] 14h-090]{90}).
%%  Otherwise, LaTeX will issue an error. 

\section{Introduction}

Classical novae result from a thermonuclear runaway (TNR) on the surface
of a white dwarf that accretes material in a close binary system
~\citep[e.g.,][]{war95, war08}.
Both Galactic and extragalactic observations have suggested that there
may exist two distinct populations
of novae~\citep[e.g.,][]{del92,del98,sha08}.
Galactic observations suggest that novae
associated with the disk were on average more luminous and faded
more quickly
than novae thought to be associated with the bulge \citep[e.g.,][]
{due90, del92}.
However, the interpretation
of Galactic nova data is complicated by
interstellar extinction, which can be significant and varies widely with
line-of-sight to a particular nova. Furthermore,
extinction hampers the discovery of a significant fraction of Galactic novae,
with only about one in four of the $\sim$35 novae that are
thought to erupt each year \citep{sha97, sha02, dar06} being
discovered and subsequently studied in any detail~\citep[e.g.,][]{hou10}.
Recent work has concentrated on
extragalactic observations which offer more promise
for understanding nova populations.
The most thoroughly studied extragalactic system has been M31,
where more than 800 novae
have been discovered over the past century \citep{pie07, pie10, sha08}.
Despite the considerable data amassed in recent years, evidence for
distinct nova populations in extragalactic systems is conflicting, with
several studies~\citep[e.g.,][]{sha00, fer03, wil04, hor08, sha11a, sha11b}
failing
to establish any definitive relationship between nova properties
and stellar population.

A promising avenue for the study of nova populations involves
the classification of nova spectra shortly after eruption.
Two decades ago \citet{wil92} established
that the spectra of Galactic novae
can be divided into one of two principal spectroscopic types, Fe~II and He/N,
based on the emission lines in their spectra.
Novae displaying prominent Fe~II emission (the ``Fe~II" novae)
usually show P Cygni absorption profiles, and evolve
more slowly, have lower expansion velocities, and lower levels of ionization,
compared to novae with strong lines of He and N (the ``He/N" novae).
The spectroscopic types are believed to be related to fundamental properties
of the progenitor binary such as the white dwarf mass.

Increased access in recent years
to queue scheduling on large telescopes,
such as the Hobby-Eberly Telescope (HET), has made it
feasible to conduct spectroscopic surveys of novae in nearby galaxies.
A comprehensive photometric and spectroscopic study of M31's
nova population has been recently published by \citet{sha11b}.
Despite the wealth of data (spectroscopic types for a total of 91 novae
were presented) there was no clear dependence of a nova's spectroscopic type
on spatial position in M31.
This result may be misleading, however, since
the relatively high inclination of M31 to the plane
of the sky ($i\sim77^{\circ}$) makes
it difficult to assign an unambiguous position within M31 to a given nova,
particularly near the apparent center of M31 where the foreground
disk is superimposed on the galactic bulge. Light curve data for many of these
novae did suggest that the more rapidly declining novae have a 
slightly more extended spatial distribution.

In this paper we report the results of a spectroscopic survey of novae
in another local group galaxy, the late-type spiral M33.
Because M33 is essentially a bulgeless galaxy, novae erupting in M33 should
be representative of a pure disk population, while the nova population
in M31, on the other hand, appears
to be bulge-dominated \citep{cia87, sha01, dar06}. Thus, a direct comparison
of the properties of M33 and M31 novae will avoid the
projection effects that plague the interpretation of the M31 results, and
hopefully better address the question of whether the spectroscopic classes
of novae vary with stellar population.

\section{Observations}

\subsection{Spectroscopy}

Spectroscopic observations were obtained
using the Low-Resolution Spectrograph
\citep{hil98} on the HET.
Initially,
we employed the $g2$ grating with a 2.0$''$ slit and the GG385 blocking filter,
covering $4275-7250$\,\AA\ at a resolution of $R\sim 650$. Later, to
obtain more coverage at longer wavelength, we opted to use
the lower resolution $g1$ grating with a
1.0$''$ slit and the GG385 blocking filter. This choice increased our
wavelength coverage
to $4150-11000$\,\AA\ while yielding a resolution of $R\sim 600$.
In practice, the useful
spectral range of the $g1$ grating is limited to
$\lambda\lessim9000$\,\AA\ where the effects of order overlap are minimal.
All HET spectra were reduced using standard IRAF\footnote{
IRAF is distributed by the National Optical Astronomy Observatory,
which is operated by the Association for Research in Astronomy, Inc.
under cooperative agreement with the National Science Foundation.}
routines to flat-field
the data and to optimally extract the spectra.
A summary of the HET observations is given in Table~\ref{hetdata}.

We obtained a total of six spectra of M33 nova candidates, which are
shown in Figures~\ref{fig1} and \ref{fig2}.
Spectra were placed on relative flux scales through comparison with
observations of spectrophotometric standards routinely used at the HET.
Because the observations were made under a variety of atmospheric
conditions with the stellar image typically overfilling the spectrograph slit,
our data cannot be considered spectrophotometric. Thus, all spectra
have been displayed on a relative flux scale.
In the caption for each figure
we have indicated the time elapsed between discovery of the nova
(not necessarily maximum light) and the date of our spectroscopy.

\subsection{Photometry}

To complement our spectroscopic survey, we were able to obtain
photometry sufficient to produce light curves of
four of the six novae in our spectroscopic survey. The data are given in
Tables~\ref{photobs1} and \ref{photobs2}, with the light curves presented
in Figure~\ref{fig3}.
Our primary motivation was to measure nova fade rates ($t_2$) that could
then be correlated with other properties, such as spectroscopic class.
The photometric data consist both
of targeted (mostly $B$ and $V\/$-band) observations,
which were obtained primarily with the
Liverpool Telescope \citep[LT,][]{ste04}
and the Faulkes Telescope North \citep[FTN,][]{bug07}.
The LT and FTN data were reduced using a combination of IRAF
and Starlink software, calibrated using standard stars from
\citet{lan92}, and checked against secondary standards
from \citet{mag92}, \citet{hai94}, and \citet{mas06}.

\section{Spectroscopic Classification of Novae}

The spectra of novae shortly after eruption (days to weeks) are characterized
by an emission-line spectrum that is dominated by Balmer lines. In addition,
novae also often display prominent emission lines
of either Fe~II (multiplet 42 is often the strongest)
or He and N in various stages
of ionization. The former group, referred to as the ``Fe~II" novae
by \citet{wil92} are often characterized by P Cygni-type line profiles,
relatively narrow line widths (FWHM H$\alpha$ typically less than
2000 km~s$^{-1}$),
along with relatively slow spectral development over timescales of weeks.
The latter class of novae, referred to as the ``He/N" novae, display
higher excitation emission lines that are generally
broader (FWHM of H$\alpha$ $\grtsim$ 2500~km~s$^{-1}$) with more rectangular,
castellated or flat-topped profiles.
It is remarkable that nova spectra can usually be
classified into one of these two distinct groups, with Fe~II novae
making up $\sim$80\% of Galactic and M31 novae \citep{sha07a, sha11b}.
A small fraction of novae appear to have
characteristics of both classes. They are referred to as either
``hybrid" or broad-lined Fe~II (Fe~IIb) novae. These novae appear
similar to Fe~II novae shortly after eruption, but the lines are
broader than those seen in a typical Fe~II nova. Later they may
evolve to display a typical He/N spectrum. On the other hand
He/N novae are never seen to evolve into Fe~II novae.
The classifications are robust, and with the exception of
the hybrid novae, they are not particularly sensitive to the precise
time during the first few weeks after eruption when the spectra
are obtained.

Nova outbursts result from a thermonuclear runaway that occurs in the
degenerate surface layers of a white dwarf that accretes matter from its
companion. The resulting eruption
ejects some or all of the accreted material, and in some cases may
dredge up material from the white dwarf itself.
Whether a particular nova becomes a Fe~II, He/N or hybrid system ultimately
must depend on the properties of the progenitor binary (principally
the white dwarf mass and its accretion rate), which govern the physical
conditions in the accreted layer at the time the nova is triggered.
Numerical models of nova eruptions
suggest that gas is ejected from the white dwarf in two distinct stages:
a discrete shell of gas ejected at the time of eruption followed by
steady mass loss in a wind \citep{wil92}. Radiation from the ejected gas
then produces the emission-line spectrum that is typical of novae
shortly after eruption. According to \citet{wil92},
the fundamental characteristics of the post-eruption
nova spectrum, and thus the spectroscopic classification,
depends on whether the dominant emission
is produced in the discrete shell or in the wind. In an He/N nova
it is thought that a relatively small amount of gas is ejected quickly, with
little contribution from a wind, causing the spectrum to be
dominated by emission from a high-velocity shell ionized by
the hot white dwarf. In an Fe~II nova, a larger accreted mass results
in a more massive ejecta consisting of both a relatively low density,
high velocity shell and an optically-thick wind driven by residual
nuclear burning on the surface of the white dwarf.

Models of nova eruptions show that a TNR is triggered when the temperature
and density at the base of the accreted envelope become sufficiently high
for nuclear burning to take place \citep{sta08}. As shown by \citet{tow05} the
amount of mass that must be accreted to trigger a TNR (the ignition mass)
is primarily a function of the white dwarf mass and temperature, with the
latter being strongly influenced by the rate of accretion onto the
white dwarf's surface. Thus, it seems plausible to expect that nova
binaries with the most massive white dwarfs and with the highest accretion
rates should have the smallest accreted masses and the
shortest recurrence times between eruptions. Further, assuming
the ejected mass is proportional to the accreted mass, such systems
should be more likely to eject their mass in a discrete shell with
little mass left over for residual burning on the white dwarf's
surface. They would then be expected to produce He/N spectra.
On the other hand, nova progenitors harboring lower mass white dwarfs
will have to accrete a greater envelope mass prior to TNR. It is these
systems that are more likely to produce a higher mass of ejected material
with a component in the form of an optically thick wind. Such systems
are expected to be characterized by Fe~II-type spectra.

\section{Novae in M33}

The first recorded M33 nova was discovered
on a plate taken by F.G. Pease on the night of 1919 December 14 \citep{hub26}.
Discoveries continued only sporadically since then with
a total of 36 novae and nova candidates
reported in M33 up through the end of 2010
\citep[e.g.,][and references therein]{pie10, wil04, del94, sha93, ros73}\footnote {\tt See http://www.mpe.mpg.de/~m31novae/opt/m33/index.php for
a compilation of positions, discovery magnitudes and dates}.
The number of novae discovered in recent years has increased dramatically
as a result of automated surveys and increased amateur astronomer activity,
with almost half of the known M33 novae being discovered in the past 15 years.
A summary of known M33 novae is presented in Table~\ref{novatable}.

\subsection{Spectroscopic Classifications}

\begin{figure}
\includegraphics[angle=0,scale=.60]{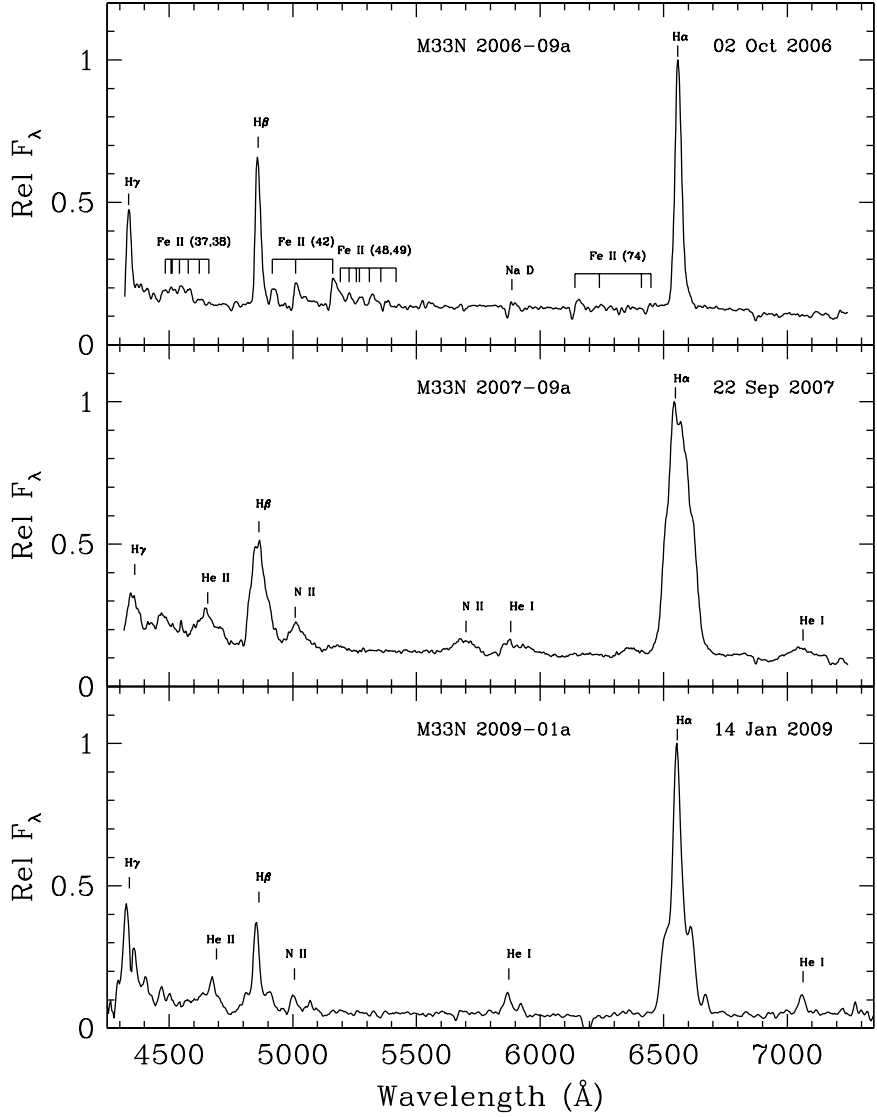}
\caption{Spectra of the M33 novae M33N~2006-09a, 2007-09a, and 2009-01a taken four, four, and seven days post-discovery, respectively. M33N~2006-09a appears to be an Fe~II system, while 2007-09a and 2009-01a are clearly He/N systems.
\label{fig1}}
\end{figure}

\begin{figure}
\includegraphics[angle=0,scale=.60]{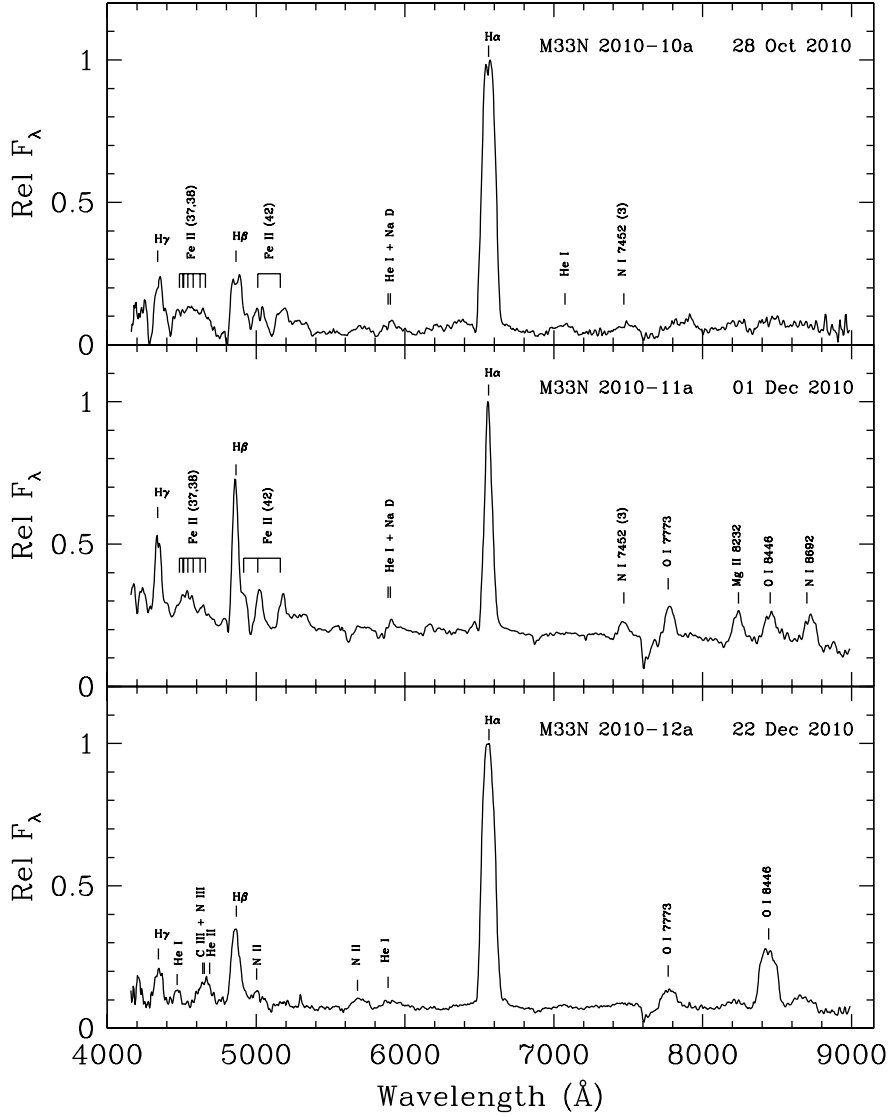}
\caption{Spectra of the M33 novae M33N~1010-10a, 2010-11a, and 2010-12a, obtained two, four, and five days post-discovery, respectively. M33N~2010-10a and 2010-11a appear to be FeIIb (or Hybrid) novae with 2010-12a belonging to the He/N class.
\label{fig2}}
\end{figure}

Given the transient nature of novae and the challenges of scheduling
time on large telescopes with short notice, it is not surprising that
the vast majority of M33 nova candidates have not been confirmed
spectroscopically. The first known spectrum was reported less than
a decade ago by \citet{sch03}
who classified M33N 2003-09a as a member of the Fe~II spectroscopic class.
As part of a spectroscopic survey of novae in local group galaxies with the HET
that began in 2006, we have obtained spectra of an additional
six M33 novae over the past five years (see Figures~\ref{fig1} and \ref{fig2}).
During this period, \citet{dim08} obtained a spectrum of 2008-02a and
concluded that the system was a member of the Fe~II class. Thus, there are
a total of eight M33 novae for which a spectroscopic classification is currently
possible. Following the classification scheme of \citet{sha11b} for M31 novae,
the six spectra included in our M33 survey
were examined and subsequently assigned to one of four possible
classes: Fe~II, He/N, hybrid (also known as broad-lined Fe~II or Fe~IIb novae),
and a potentially new class of narrow-lined He/N systems, the He/Nn novae.
Below we summarize the properties of M33 novae with measured spectra.

\noindent
{\bf $\bullet$ M33N 2003-09a:}
M33N 2003-09a was discovered at Lick Observatory by M. Ganeshalinam and
W. Li with the Katzman Automated Imaging Telescope on Sep. 01.4 UT at
$m = 16.9$~\citep{gan03}. A little less than two days later
on Sep. 03.05~\citet{shp03} found the brightness of the nova
relatively unchanged
at $V=16.9$. A spectrum obtained approximately two weeks post-discovery
by \citet{sch03} with the MMT revealed the object to be a likely
member of the Fe~II spectroscopic class. The rather large reported
width of the H$\alpha$ line (FWZI $\sim5400$ km~s$^{-1}$) suggests that
the nova may possibly be a member of the Fe~IIb or hybrid class.
No information is available concerning the speed class of this nova.

\begin{figure}
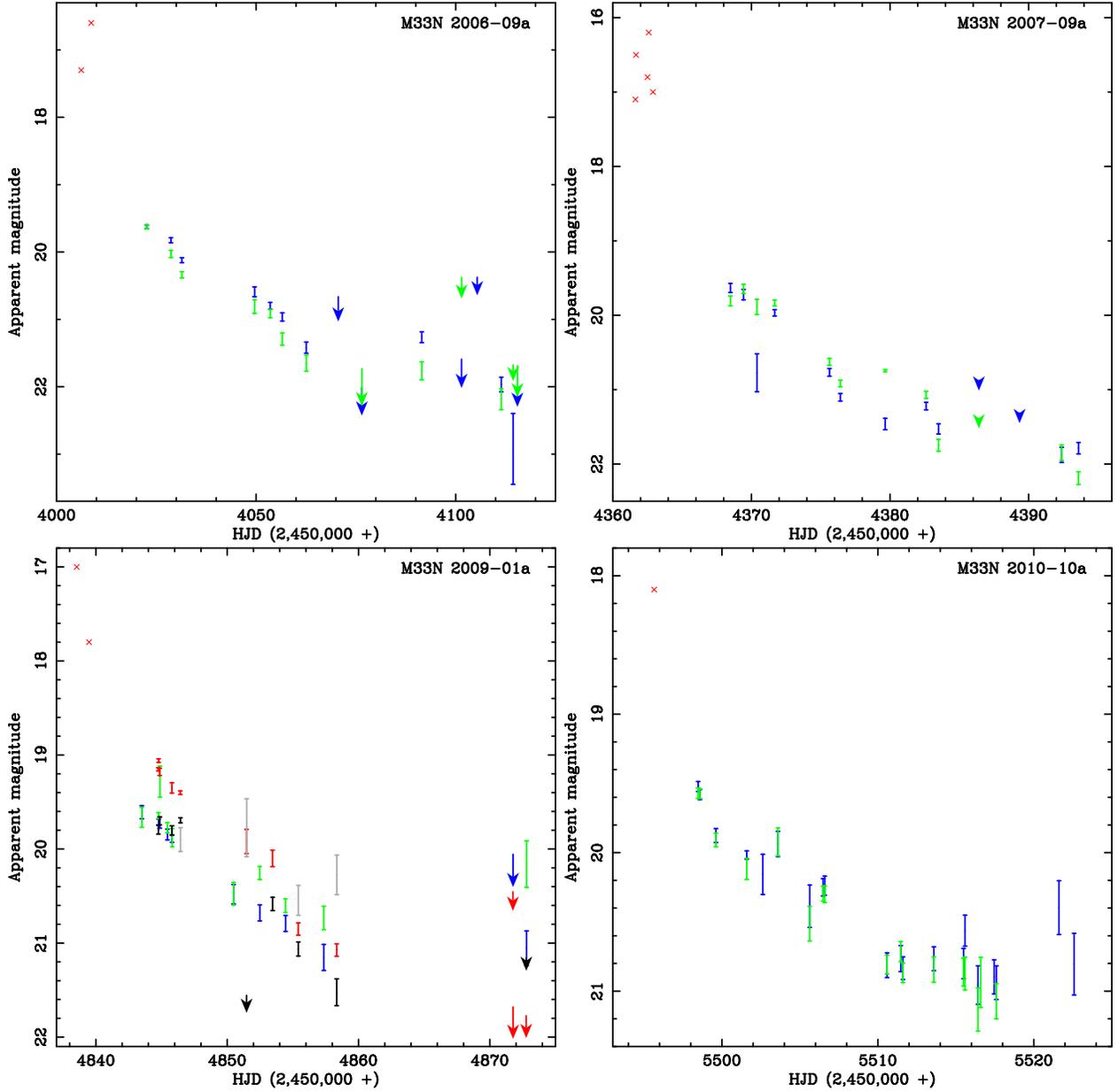

\includegraphics[angle=0,width=.50\textwidth]{f3a.eps}
\includegraphics[angle=0,width=.50\textwidth]{f3b.eps}
\newline
\includegraphics[angle=0,width=.50\textwidth]{f3c.eps}
\includegraphics[angle=0,width=.50\textwidth]{f3d.eps}
\caption{Light curves for M33N~2006-09a, M33N~2007-09a,
M33N~2009-01a, and M33N~2010-10a. Uncertainties in
measurements are shown as vertical bars with the following
colors representing the different bandpasses:
$B$ -- blue;
$V$ -- green;
$R$ -- red.
%$r'$ -- red;
%$i'$ -- black;
%$z'$ -- light grey.
Upper flux limits are indicated by downward facing arrows.
\label{fig3}}
\end{figure}

\noindent
{\bf $\bullet$ M33N 2006-09a:}
M33N 2006-09a was discovered independently by R. Quimby et al. and by
S. Nakano on Sep 28.20 UT ($m\sim16.9$) and
Sep. 30.68 UT ($m\sim16.6$), respectively~\citep{qui06,ita06}. As part
of our survey, a spectrum of 2006-09a was obtained on Oct. 02.42
with the HET~\citep{sha06}.
The spectrum, shown in Figure~\ref{fig1}, reveals the nova
to be a typical member of the Fe~II spectroscopic class.
The light curve, shown in Figure~\ref{fig3}, indicates that the nova faded
moderately slowly with $t_2[B] \sim t_2[V] \sim 34$~days.

\noindent
{\bf $\bullet$ M33N 2007-09a:}
M33N 2007-09a was discovered by K. Nishiyama and F. Kabashima on
Sep. 18.63 UT at $m \sim17.1$~\citep{nak07}. A spectrum obtained
approximately two days post-discovery by \citet{wag07} revealed
intense and broad Balmer and He~I emission lines indicating
that the nova was a member of the He/N class. Our HET spectrum, which
was obtained approximately four days post-discovery on Sep. 22.25 UT
\citep{sha07b},
confirms the He/N classification (see Figure~\ref{fig1}). Subsequent
photometry obtained with the LT revealed
that the nova faded relatively rapidly
(consistent with the He/N classification)
with $t_2[B] \sim 11$~d and $t_2[V] \sim 6$~d (see Figure~\ref{fig3}).

\noindent
{\bf $\bullet$ M33N 2008-02a:}
K. Nishiyama and F. Kabashima discovered M33N 2008-02a
on Feb. 27.47 UT at $m \sim16.5$~\citep{nak08}. Subsequently,
a spectrum obtained on Mar. 2.80 UT revealed the nova
to be a member of the Fe~II class~\citep{dim08}.
No light curve information is available for this nova.

\noindent
{\bf $\bullet$ M33N 2009-01a:}
M33N 2009-01a was discovered by
K. Nishiyama and F. Kabashima, who found the nova to reach
$m \sim17.0$ on Jan 07.54 UT~\citep{nak09}.
We obtained an HET spectrum of the nova
approximately a week post-discovery on Jan. 14.14~UT~\citep{sha09}.
The spectrum, shown in Figure~\ref{fig1}, reveals H, He, and N emission
lines with a complex structure consisting of both a broad base component
with a narrower core component (see Table~\ref{balmerline}).
The light curve obtained from our LT photometry and shown in Figure~\ref{fig3}
reveals that the nova was moderately fast as expected for
an He/N nova, being characterized by $t_2 \sim 17$~days and $t_2 \sim 12$~days
for the $B$ and $V$ bandpasses, respectively.

\noindent
{\bf $\bullet$ M33N 2010-10a:}
Like the previous 3 M33 novae, M33N 2010-10a was also discovered
by K. Nishiyama and F. Kabashima who found the nova at $m \sim18.1$
(unfiltered) on Oct. 26.654 UT~\citep{yus10a}. In order to
classify the nova, we obtained an
HET spectrum on Oct. 28.37 UT (see Figure~\ref{fig2}), which revealed
broad Balmer (FWHM H$\alpha$ $\sim4200$~km~s$^{-1}$), He, N, and Fe~II emission
lines~\citep{sha10a}.
The presence of the Fe~II emission suggests that this nova should
be classified as a member of the Fe~IIb, or hybrid spectroscopic class.
Photometry obtained with the LT has enabled us to produce the light curve
shown in Figure~\ref{fig3}. The decline from maximum light was moderately fast
with $t_2 \sim 23$~d and $t_2 \sim 20$~d, for the $B$ and $V$ bandpasses,
respectively.

\noindent
{\bf $\bullet$ M33N 2010-11a:}
M33N 2010-11a was discovered by J. Ruan and X. Gao on Nov. 27.53 UT at
$m = 18.6$ and independently by K. Nishiyama on Nov. 28.54 UT at
$m = 16.7$. The nova continued to brighten, reaching $m = 16.1$ on
Nov 29.064 UT~\citep{yus10b}. We obtained a spectrum of 2010-11a
(see Figure~\ref{fig2}) on Dec. 01.05~UT~\citep{sha10b}.
The spectrum is characterized by
relatively broad Balmer, He I, Fe~II (and possibly N I) emission lines
(FWHM H$\alpha$ $\sim2600$~km~s$^{-1}$), and can best be described
as that of a broad-lined Fe~II, or hybrid nova.

\noindent
{\bf $\bullet$ M33N 2010-12a:}
M33N 2010-12a was discovered on Dec. 17.42 UT
at $m = 16.6$~\citep{yus10c}. We obtained a spectrum (see Figure~\ref{fig2})
of the nova 5 days later on Dec. 22.20~UT with the HET~\citep{sha10c}.
The broad Balmer, He, and N emission (FWHM H$\alpha$ $\sim4100$~km~s$^{-1}$)
clearly establish the nova as a member of the He/N spectroscopic class.

In summary, when all eight novae with observed spectra
are considered (see Table~\ref{novatable}),
we find that five of the eight systems are either He/N or related systems
(Fe~IIb and hybrid), with
Fe~II novae making up less than 40\% of the total. Despite the
relatively small number of M33 novae that have been classified, this result
appears to be in sharp contrast to the data for M31 and the Galaxy where
Fe~II novae comprise roughly 80\% and 70\% of the total, respectively
\citep{sha11b}.
The somewhat higher percentage of Fe~II novae observed in M31
relative to the Galaxy
may reflect the fact that nova surveys have concentrated primarily on the
bulge of M31, whereas Galactic data are biased to relatively
nearby novae mostly located in the Galactic disk.

It is possible that as a result of outburst evolution our spectroscopic
classifications may depend on the precise timing of the observations.
For example, in hybrid novae, as the outburst progresses the spectrum
evolves from a (broad-lined) Fe~II spectrum to that resembling an He/N
spectrum. On the other hand, novae have not been observed to evolve
in the opposite sense, from He/N to Fe~II class. Thus, if our
classifications do evolve, it will likely result in fewer novae being
classified as Fe~II, and will exacerbate the discrepancy with
the M31 and Galactic data.

\subsubsection{Expansion Velocities}

One of the defining properties of the He/N spectroscopic class
is that the emission line widths are considerably broader than
those seen in the Fe~II novae.
Specifically, \citet{wil92} found that the emission lines of Galactic
novae in the He/N class are typically characterized by a half-width at zero
intensity, HWZI $>2500$ km~s$^{-1}$. Empirically,
we have found that for most nova line profiles the HWZI~$\simeq$~FWHM;
since the latter is the more
easily measured quantity, we have followed
\citet{sha11b} and adopted the FWHM to characterize
the spectra in our survey.
The values of the FWHM and the equivalent widths of H$\alpha$
and H$\beta$ in our nova spectra are given in Table~\ref{balmerline}.
Without exception, as in M31 \citep{sha11b},
the novae belonging to the He/N class are characterized by H$\alpha$
FWHM~$>2500$~km~s$^{-1}$, while the Fe~II systems all have
an FWHM less than this value.

Although the emission line width is expected to be correlated
with the expansion velocity
of the nova ejecta, the FWHM does not necessarily yield
the expansion velocity directly.
In an Fe~II nova,
the lines are mainly produced in a wind, which originates
at a distance above the surface of the white dwarf that varies
as the outburst evolves. Thus, the escape velocity for this wind
is smaller than that
at the white dwarf's surface. As a result, the derived
expansion velocity (and hence
line width) may decrease with the time elapsed
since eruption. In an He/N nova, on the other hand,
the broad emission features are
believed to be formed mainly in a
discrete, optically-thin shell ejected at relatively high velocity
from near the white dwarf's surface.
The line profiles are expected to be flat-topped
with the FWHM closely approximating the ejection velocity
of the shell.

\subsection{Light Curve Properties}

To further explore the properties of the novae in our survey,
whenever possible we have augmented our spectroscopic data with
available photometric observations. Unfortunately, we could find no light curve
information for M33 novae erupting prior to the start of our
HET spectroscopic survey in 2006.
Nevertheless,
we have sufficient photometric data to estimate decline rates
for half of the novae (four of eight) in our spectroscopic sample.

A convenient
and widely used parameterization of the decline rate
is $t_2$, which represents the time (in days) for a nova
to decline 2 mag from maximum light. According to the criteria
of \citet{war08}, novae
with $t_2\lessim25$ days are considered ``fast" or ``very fast",
with the slowest novae characterized by $t_2$ values of
several months or longer.
Rates of decline, and corresponding values of $t_2$, have been measured
for the 4 novae in our photometric sample
by performing weighted linear least-squares fits to the declining portion
of the light curves that extend up to 3 mag below peak.
In an attempt to account for systematic errors in the individual
photometric measurements, the weights used in the fits were composed
of the sum of the formal errors on the individual photometric
measurements plus a constant systematic error estimate of 0.1 mag.
The net effect of including the systematic error component was a
reduction of the relative weighting of points with small formal errors and a
corresponding increase in the formal errors of the best-fit parameters
and in the uncertainties in $t_2$ derived from them.

Because our photometric observations do not always
begin immediately after discovery, and the date of discovery does not
always represent the date of eruption,
we have made two modifications to our photometric data in order to
better estimate the light curve parameters.
First, when available,
we have augmented our light curve data with
the discovery dates and magnitudes given in the catalog
of \citet{pie10}\footnote{\tt see also http://www.mpe.mpg.de/\~m31novae/opt/m33/index.php}.
Secondly, for some novae we have
modified (brightened) the peak magnitude slightly
through an extrapolation of the declining portion
of the light curve by up to 2.5 days pre-discovery in cases where
upper flux limits (within five days of discovery) are available.
Finally, apparent magnitudes at the time of discovery have been converted to
estimates of the absolute magnitude at maximum light by adopting 
distance moduli for M33 of $\mu_B = 25.35$ and $\mu_V = 25.26$
\citep{pel11}.
The light curve parameters resulting from our analysis
are given in Table~\ref{lcparam}.
In agreement with the results of both the Galactic study by~\citet{del98}
and the M31 study by \citet{sha11b},
it appears that the He/N novae in M33 are generally ``faster" than
their Fe~II counterparts, as expected for novae with more massive
white dwarfs~\citep[e.g.,][]{liv92}.

\subsection{The Spatial Distribution of M33 Novae}

\begin{figure}
\includegraphics[angle=-90,scale=.85]{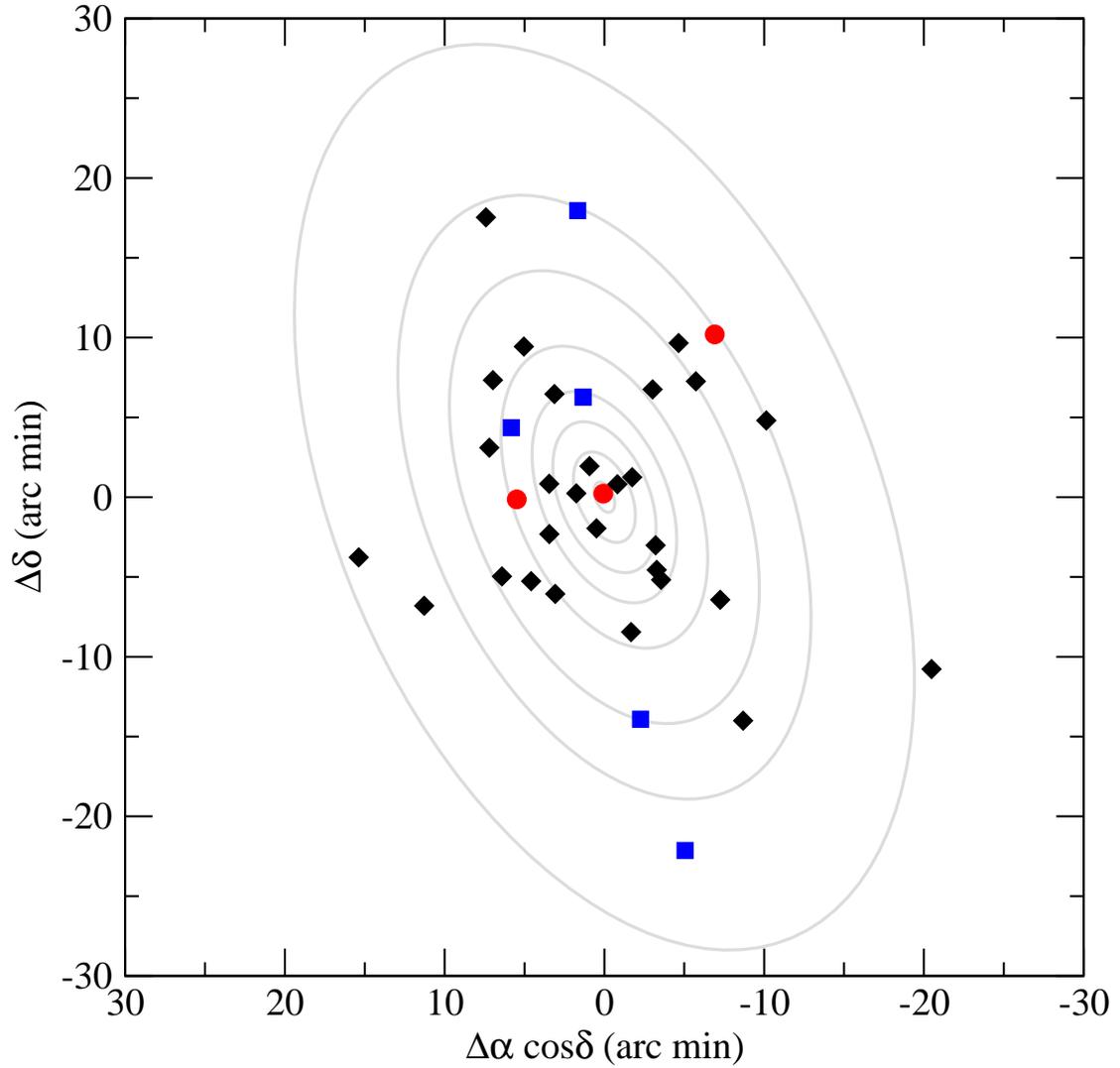}
\caption{Spatial distribution of the 36 recorded novae in M33.
The Fe~II novae are
indicated by filled red circles, while the
He/N and Fe~IIb (hybrid) novae are represented by filled
blue squares. The black diamonds represent the 28 novae
with unknown spectroscopic types. The gray ellipses
represent elliptical isophotes from the surface photometry of
\citet{reg94}.
\label{fig4}}
\end{figure}

The projected positions of the 36 known M33 novae from Table~\ref{novatable}
are shown in Figure~\ref{fig4}. For the 8 novae with known spectroscopic class
we have plotted the Fe~II systems as filled circles and
the He/N and Hybrid novae as filled squares.
Observations by \citet{del98} suggest that Fe~II and He/N novae are
associated with different stellar populations: the He/N novae
primarily with the Galactic disk and the Fe~II novae with thick disk and bulge.
Given that M33 is a nearly bulgeless galaxy,
classified as an Scd galaxy \citep{tul88},
the spatial distributions of the Fe~II and He/N novae are not expected
to differ significantly.

\begin{figure}
\includegraphics[angle=-90,scale=.65]{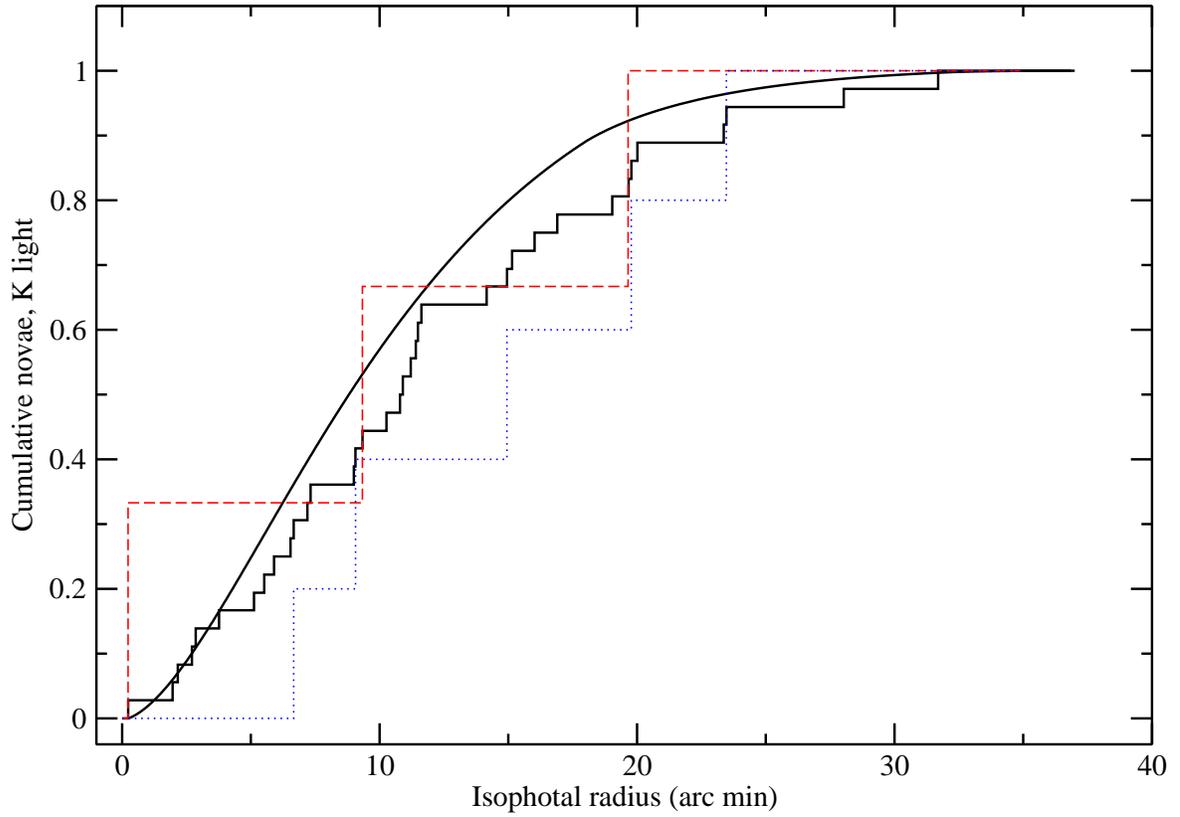}
\caption{Cumulative distribution of the M33 novae compared with
background K light from \citet{reg94}.
A K-S test indicates a 39\% probability that the distributions
would differ by more than they do under the null hypothesis that the
distributions are drawn from the same parent population.
For comparison, the Fe~II systems
(red dashed line) are compared with the He/N (Fe~IIb) systems (blue
dotted line).
\label{fig5}}
\end{figure}

M33 is oriented at an angle of $57^{\circ}$ with respect to the plane
of the sky. Thus, in order to 
approximate the true position of a nova within M33,
we have assigned each nova an isophotal radius, defined as
the length of the semi-major axis of an elliptical
isophote computed from the $R$-band surface photometry of \citet{ken87}
that passes through the observed position of the nova.
In Figure~\ref{fig5}, we show the
cumulative distribution the M33 novae compared with the cumulative
$K$-band light from the surface photometry of \citet{reg94}.
A Kolmogorov-Smirnov (K-S) test reveals that the distributions
would be expected to differ by more than that
observed 39\% of the time if they were drawn
from the same parent population. Thus, there is no reason to
reject the hypothesis that the novae follow
the light distribution in M33. We have also shown the cumulative distributions
of the Fe~II and He/N (and hybrid) novae separately,
although we have insufficient data to draw any conclusions regarding
whether these distributions may differ.

\section{Discussion and Conclusions}

The answer to the
question of whether or not there exist two distinct populations of novae
has remained elusive, with observational support for both
sides of the issue. For example, an analysis of spectroscopic observations
of Galactic novae by \citet{del98} suggests that He/N novae are generally
located closer to the Galactic plane than are the Fe~II novae,
suggesting that the He/N novae belong to a younger stellar population
compared with the latter class of novae. On the other hand, the recent
spectroscopic survey of novae in M31 by \cite{sha11b} finds no
significant difference in the spatial distributions of the He/N and Fe~II
novae in that galaxy.

In an attempt to gain further insight into the question of whether
the spectroscopic class of novae is sensitive to stellar population,
we have initiated a spectroscopic survey of novae in the late-type,
nearly bulgeless spiral galaxy, M33. Despite the fact that
the absolute nova rate
in M33 is only $\sim2.5$ per year \citep{wil04}, we were able to secure
spectra for six novae thus far since our survey began in 2006. After including
spectroscopic observations of two additional novae from the literature,
we were ultimately able to
establish spectroscopic classes for a total of eight novae in M33.
Of the eight, only two novae (M33N 2006-09a and 2008-02a)
are clearly members of the Fe~II class, with a third nova,
2003-09a, possibly being a member of the Fe~IIb class.
Of the remaining novae, three are clearly He/N novae, with two
being Fe~IIb or hybrid objects.

Although the number of spectra available for M33 is currently small relative
to M31, it is already clear that the fraction of Fe II novae
in M33 is surprisingly low.  We can estimate the significance of this result
as follows.  Let $p({\rm Fe~II})$ be the fraction of novae occurring within a
galaxy with the spectroscopic classification Fe II, and let
$q({\rm He/N}) = 1 - p({\rm Fe~II})$
be the fraction of novae with spectroscopic classification He/N (+ Fe IIb).
The probability of observing $N$ Fe II novae out of a sample of $M$ objects
is simply
\begin{equation}
P_{N,M} = {M! \over N! (M - N)!} \  p({\rm Fe~II})^N \, q({\rm He/N})^{M-N},
\end{equation}
and the probability of observing $N$ or fewer Fe II novae is
\begin{equation}
P_{\leq N,M} = \sum_{n=0}^N P_{n,M}.
\end{equation}
Under the null hypothesis that the novae in M33 have the same ratio of
spectroscopic types as that seen in M31, we have $p({\rm Fe~II}) = 0.8$ and
$q({\rm He/N}) = 0.2$, and the probability of observing three or fewer Fe~II
novae is $P_{\leq 3,8} = 0.0104$.
In other words, the mix of spectroscopic nova types
in M33 differs from that of M31 at the 99\% confidence level.

This result is perhaps not surprising given that
the spectroscopic class of novae is expected to depend on fundamental
physical properties of the nova progenitor binary such as
the mass of the white dwarf. The average mass of the white dwarfs in
nova binaries, in turn, is thought to be sensitive to the age of the
underlying stellar population in the sense that a younger stellar population,
like that which dominates in M33, should contain, on average, higher mass
white dwarfs~\citep[e.g.,][]{dek92,tut95,pol96}.
It therefore appears plausible that the observed difference in the mix
of nova types in M31 and M33 might be related to
the difference in stellar population between the two galaxies.

In addition to obtaining spectroscopic observations of M33 novae,
we were able to measure light curves (and resulting $t_2$ times)
for the majority of the novae
in our sample, and for half of the novae with known
spectroscopic class. Although our small sample does not allow us to
study the spatial distribution of speed class as we were able to
do in the case of M31~\citep{sha11b}, we did find, as expected,
that of the four novae with measured $t_2$, the
fastest declining systems (M33N 2007-09a and 2009-01a) were
members of the He/N
spectroscopic class, while the slowest nova, 2006-09a, was found to be
a member of the Fe~II class.

Finally, by considering the positions of all 36 novae seen to erupt in M33
over the past century~\citep{pie10}, we were able to explore
the spatial distribution of novae across the galaxy. Although the available
dataset is too limited to study the distributions of the Fe~II and He/N
novae separately, we did find that the overall nova distribution is consistent
with that expected if the nova rate is proportional to
the surface brightness distribution in the galaxy.

In order to confirm our preliminary findings,
future observational efforts should focus not only
on continued optical imaging to discover additional M33 nova candidates,
but also on the timely spectroscopic and photometric follow-up observations
required to increase the number of available spectroscopic and
speed classifications.
In addition, X-ray observations to measure the timing and duration
of the nova supersoft stage can be used provide useful constraints
on the properties of nova binaries in M33, as they have in studies
of novae in M31~\citep[e.g.][]{pie07,bod09,hen10,pie11,hen11}.

\newpage
\acknowledgments

The work presented here was made possible through
observations obtained from
facilities based throughout the world. Spectroscopic
observations were obtained with the Lick Observatory's Shane 3~m
telescope operated by the University of California and with the Marcario
Low Resolution Spectrograph on the Hobby-Eberly Telescope,
which is operated by McDonald Observatory on behalf of the University
of Texas at Austin, the Pennsylvania State University,
Stanford University, the Ludwig-Maximillians-Universitaet,
Munich, and the George-August-Universitaet, Goettingen.
Photometric observations were made using
the Liverpool Telescope, which is operated on the island of
La Palma by Liverpool John Moores University (LJMU)
in the Spanish Observatorio del
Roque de los Muchachos of the Instituto de Astrofisica de Canarias with
financial support from the UK Science and Technology Facilities Council
(STFC). Faulkes Telescope North (FTN)
is operated by the Las Cumbres Observatory Global Telescope
network. Data from FTN were obtained as part of a joint programme
between Las Cumbres Observatory and LJMU Astrophysics Research
Institute.
Finally, we thank K. Nishiyama and F. Kabashima for their tireless
efforts to monitor M33 for erupting novae, and
two anonymous referees for their comments on earlier versions
of the manuscript.
A.W.S. is also grateful
to the NSF for financial support through grants AST-0607682 and AST-1009566.

%% Use the figure environment and \plotone or \plottwo to include
%% figures and captions in your electronic submission.
%% To embed the sample graphics in
%% the file, uncomment the \plotone, \plottwo, and
%% \includegraphics commands
%%
%% If you need a layout that cannot be achieved with \plotone or
%% \plottwo, you can invoke the graphicx package directly with the
%% \includegraphics command or use \plotfiddle. For more information,
%% please see the tutorial on "Using Electronic Art with AASTeX" in the
%% documentation section at the AASTeX Web site,
%% http://www.journals.uchicago.edu/AAS/AASTeX.
%%
%% The examples below also include sample markup for submission of
%% supplemental electronic materials. As always, be sure to check
%% the instructions to authors for the journal you are submitting to
%% for specific submissions guidelines as they vary from
%% journal to journal.

%% This example uses \plotone to include an EPS file scaled to
%% 80% of its natural size with \epsscale. Its caption
%% has been written to indicate that additional figure parts will be
%% available in the electronic journal.

%% Here we use \plottwo to present two versions of the same figure,
%% one in black and white for print the other in RGB color
%% for online presentation. Note that the caption indicates
%% that a color version of the figure will be available online.
%%

\clearpage

\clearpage

\begin{deluxetable}{lcccccc}
\tabletypesize{\scriptsize}
\tablenum{1}
\tablewidth{0pt}
\tablecolumns{7}
\tablecaption{Summary of HET Spectroscopic Observations\label{hetdata}}
\tablehead{\colhead{} & \colhead{R.A.} & \colhead{Decl.} & \colhead{}&\colhead{Exp.}
 & \colhead{Coverage} & \colhead{}  \\
\colhead{Nova} & \colhead{(2000.0)} & \colhead{(2000.0)} & \colhead{UT Date} & \colhead{(sec)} & \colhead{(\AA)} & \colhead{Weather}}
\startdata
M33N 2006-09a  &  01 33 18.7  & 30 49 49 & 02 Oct 2006 &1200 &4300--7300 & spec \\
M33N 2007-09a  &  01 33 58.6  & 30 57 34 & 22 Sep 2007 &1200 &4300--7300 & spec \\
M33N 2009-01a  &  01 33 40.4  & 30 25 42 & 14 Jan 2009 &1200 &4300--7300 & phot \\
M33N 2010-10a  &  01 33 57.1  & 30 45 53 & 28 Oct 2010 &1300 &4150--9000 & phot \\
M33N 2010-11a  &  01 33 47.5  & 30 17 28 & 01 Dec 2010 &1000 &4150--9000 & spec \\
M33N 2010-12a  &  01 34 18.0  & 30 43 58 & 22 Dec 2010 &1200 &4150--9000 & spec \\
\enddata
\end{deluxetable}

\clearpage

\begin{planotable}{lcr}
\tabletypesize{\scriptsize}
\tablenum{2}
\tablewidth{0pt}
\tablecolumns{3}
\tablecaption{Photometric Observations\tablenotemark{a}\label{photobs1}}
\tablehead{\colhead{$\mathrm{MJD}$} & \colhead{} & \colhead{} \\ \colhead{($50,000+$)} & \colhead{Filter} & \colhead{Mag}}

\startdata

\cutinhead{M33N 2006-09a}

4022.119 & $B$ & $19.626\pm0.028$\\
4028.185 & $B$ & $19.825\pm0.037$\\
4030.918 & $B$ & $20.122\pm0.036$\\
4049.125 & $B$ & $20.591\pm0.073$\\
4053.045 & $B$ & $20.799\pm0.051$\\
4056.056 & $B$ & $20.964\pm0.061$\\
4062.083 & $B$ & $21.420\pm0.083$\\
4070.081 & $B$ & $>20.663\pm0.368$\\
4075.991 & $B$ & $>22.011\pm0.417$\\
4090.996 & $B$ & $21.265\pm0.081$\\
4100.974 & $B$ & $>21.591\pm0.427$\\
4104.896 & $B$ & $>20.375\pm0.268$\\
4110.916 & $B$ & $21.967\pm0.108$\\
4113.898 & $B$ & $22.923\pm0.526$\\
4114.964 & $B$ & $>22.110\pm0.189$\\

4022.122 & $V$ & $19.629\pm0.029$\\
4028.188 & $V$ & $20.029\pm0.053$\\
4030.921 & $V$ & $20.339\pm0.046$\\
4049.128 & $V$ & $20.810\pm0.101$\\
4053.049 & $V$ & $20.914\pm0.061$\\
4056.059 & $V$ & $21.292\pm0.091$\\
4062.086 & $V$ & $21.650\pm0.119$\\
4075.995 & $V$ & $>21.729\pm0.560$\\
4090.999 & $V$ & $21.763\pm0.134$\\
4100.978 & $V$ & $>20.377\pm0.315$\\
4110.919 & $V$ & $22.185\pm0.156$\\
4113.901 & $V$ & $>21.675\pm0.242$\\
4114.967 & $V$ & $>21.688\pm0.469$\\

\cutinhead{M33N 2007-09a}

4368.986 & $B$ & $19.807\pm0.066$\\
4369.931 & $B$ & $19.647\pm0.061$\\
4370.896 & $B$ & $19.889\pm0.102$\\
4372.176 & $B$ & $19.835\pm0.039$\\
4376.130 & $B$ & $20.625\pm0.046$\\
4376.919 & $B$ & $20.917\pm0.045$\\
4380.152 & $B$ & $20.745\pm0.015$\\
4383.104 & $B$ & $21.070\pm0.048$\\
4384.000 & $B$ & $21.748\pm0.080$\\
4386.909 & $B$ & $>21.443\pm0.078$\\
4392.885 & $B$ & $21.847\pm0.104$\\
4394.099 & $B$ & $22.189\pm0.086$\\

4368.989 & $V$ & $19.633\pm0.061$\\
4369.934 & $V$ & $19.723\pm0.070$\\
4370.899 & $V$ & $20.773\pm0.255$\\
4372.179 & $V$ & $19.967\pm0.042$\\
4376.132 & $V$ & $20.768\pm0.052$\\
4376.921 & $V$ & $21.103\pm0.051$\\
4380.154 & $V$ & $21.462\pm0.077$\\
4383.107 & $V$ & $21.221\pm0.051$\\
4384.003 & $V$ & $21.528\pm0.069$\\
4386.911 & $V$ & $>20.951\pm0.061$\\
4389.848 & $V$ & $>21.372\pm0.078$\\
4392.888 & $V$ & $21.877\pm0.102$\\
4394.102 & $V$ & $21.787\pm0.076$\\

\cutinhead{M33N 2009-01a}

4843.999 & $B$ & $19.663\pm0.106$\\
4845.944 & $B$ & $19.773\pm0.054$\\
4850.981 & $B$ & $20.476\pm0.120$\\
4852.971 & $B$ & $20.255\pm0.069$\\
4854.932 & $B$ & $20.602\pm0.073$\\
4857.860 & $B$ & $20.733\pm0.124$\\

4844.266 & $B$ & $19.713\pm0.030$\tablenotemark{b}\\      
4844.374 & $B$ & $19.719\pm0.059$\tablenotemark{b}\\
4845.304 & $B$ & $19.891\pm0.038$\tablenotemark{b}\\
4871.271 & $B$ & $>20.056\pm0.349$\tablenotemark{b}\\
4872.299 & $B$ & $>21.055\pm0.184$\tablenotemark{b}\\

4844.001 & $V$ & $19.608\pm0.070$\\
4845.948 & $V$ & $19.846\pm0.057$\\
4850.984 & $V$ & $20.481\pm0.103$\\
4852.982 & $V$ & $20.679\pm0.085$\\
4854.944 & $V$ & $20.791\pm0.085$\\
4857.862 & $V$ & $21.153\pm0.139$\\

4844.271 & $V$  &$19.645\pm0.032$\tablenotemark{b}\\
4844.379 & $V$  &$19.283\pm0.165$\tablenotemark{b}\\
4845.309 & $V$  &$19.879\pm0.097$\tablenotemark{b}\\
4872.301 & $V$  &$>20.160\pm0.247$\tablenotemark{b}\\

4845.934 & $r'$ &$19.401\pm0.019$\\
4850.969 & $r'$ &$19.920\pm0.129$\\
4852.951 & $r'$ &$20.100\pm0.088$\\
4854.912 & $r'$ &$20.851\pm0.065$\\
4857.840 & $r'$ &$21.074\pm0.066$\\
        
4844.251 & $r'$ &$19.152\pm0.017$\tablenotemark{b}\\
4844.283 & $r'$ &$19.060\pm0.019$\tablenotemark{b}\\
4844.359 & $r'$ &$19.180\pm0.038$\tablenotemark{b}\\
4845.289 & $r'$ &$19.350\pm0.055$\tablenotemark{b}\\
4871.256 & $r'$ &$>20.455\pm0.202$\tablenotemark{b}\\
4871.274 & $r'$ &$>21.679\pm0.340$\tablenotemark{b}\\
4872.270 & $r'$ &$>21.772\pm0.238$\tablenotemark{b}\\

4845.938 & $i'$ &$19.695\pm0.027$\\
4850.973 & $i'$ &$>21.556\pm0.186$\\
4852.955 & $i'$ &$20.584\pm0.071$\\
4854.916 & $i'$ &$21.063\pm0.075$\\
4857.843 & $i'$ &$21.523\pm0.143$\\
                                                             
4844.256 & $i'$ &$19.795\pm0.047$\tablenotemark{b}\\
4844.364 & $i'$ &$19.700\pm0.043$\tablenotemark{b}\\
4845.294 & $i'$ &$19.802\pm0.049$\tablenotemark{b}\\
4872.275 & $i'$ &$>21.158\pm0.137$\tablenotemark{b}\\

4845.943 & $z'$ &$19.900\pm0.126$\\
4850.977 & $z'$ &$19.774\pm0.308$\\
4854.920 & $z'$ &$20.546\pm0.159$\\
4857.848 & $z'$ &$20.275\pm0.210$\\

\cutinhead{M33N 2010-10a}

5499.084 & $B$ & $19.574\pm0.035$\tablenotemark{c}\\
5498.972 & $B$ & $19.570\pm0.038$\tablenotemark{c}\\
5500.112 & $B$ & $19.908\pm0.049$\tablenotemark{c}\\
5502.088 & $B$ & $20.117\pm0.076$\tablenotemark{c}\\
5504.087 & $B$ & $19.922\pm0.101$\tablenotemark{c}\\
5506.126 & $B$ & $20.513\pm0.125$\tablenotemark{c}\\
5507.093 & $B$ & $20.300\pm0.060$\tablenotemark{c}\\
5506.984 & $B$ & $20.296\pm0.054$\tablenotemark{c}\\
5511.085 & $B$ & $20.808\pm0.068$\tablenotemark{c}\\
5511.965 & $B$ & $20.714\pm0.072$\tablenotemark{c}\\
5512.106 & $B$ & $20.868\pm0.070$\tablenotemark{c}\\
5514.087 & $B$ & $20.844\pm0.092$\tablenotemark{c}\\
5516.084 & $B$ & $20.873\pm0.119$\tablenotemark{c}\\
5515.990 & $B$ & $20.862\pm0.101$\tablenotemark{c}\\
5517.094 & $B$ & $20.937\pm0.181$\tablenotemark{c}\\
5516.918 & $B$ & $21.133\pm0.156$\tablenotemark{c}\\
5518.095 & $B$ & $21.073\pm0.127$\tablenotemark{c}\\

5499.087 & $V$ & $19.580\pm0.036$\\
5498.975 & $V$ & $19.522\pm0.036$\\
5500.115 & $V$ & $19.876\pm0.050$\\
5502.091 & $V$ & $20.017\pm0.030$\\
5503.121 & $V$ & $20.157\pm0.145$\\
5504.090 & $V$ & $19.937\pm0.091$\\
5506.129 & $V$ & $20.387\pm0.153$\\
5507.095 & $V$ & $20.238\pm0.069$\\
5506.986 & $V$ & $20.250\pm0.063$\\
5511.088 & $V$ & $20.813\pm0.089$\\
5511.968 & $V$ & $20.765\pm0.094$\\
5512.109 & $V$ & $20.834\pm0.082$\\
5514.090 & $V$ & $20.766\pm0.086$\\
5516.087 & $V$ & $20.563\pm0.112$\tablenotemark{c}\\
5515.993 & $V$ & $20.801\pm0.109$\tablenotemark{c}\\
5516.921 & $V$ & $20.956\pm0.139$\tablenotemark{c}\\
5518.097 & $V$ & $20.939\pm0.122$\tablenotemark{c}\\
5517.955 & $V$ & $20.897\pm0.123$\tablenotemark{c}\\
5522.114 & $V$ & $20.397\pm0.194$\tablenotemark{c}\\
5523.086 & $V$ & $20.805\pm0.223$\tablenotemark{c}\\

\enddata
\tablenotetext{a}{Data from Liverpool Telescope unless otherwise noted.}
\tablenotetext{b}{Data from Faulkes Telescope North.}
\tablenotetext{c}{Photometry dominated by near neighbor - J013357.15$+$304551.6 $V=20.785$, $B=21.148$.}
\end{planotable}

\clearpage

\begin{planotable}{lccc}
\tabletypesize{\scriptsize}
\tablenum{3}
\tablewidth{0pt}
\tablecolumns{4}
\tablecaption{Supplemental Photometry\label{photobs2}}
\tablehead{\colhead{$\mathrm{MJD}$} & \colhead{} & \colhead{} & \colhead{} \\ \colhead{($50,000+$)} & \colhead{Filter} & \colhead{Mag} & \colhead{References\tablenotemark{a}}}

\startdata

\cutinhead{M33N 2006-09a}

4006.200 & $W$ & $17.3$ & 1\\
4008.684 & $W$ & $17.6$ & 2\\

\cutinhead{M33N 2007-09a}

4361.630 & $W$ & $17.1$ & 3\\ 
4361.677 & $W$ & $16.5$ & 3\\
4362.501 & $R$ & $16.2$ & 3\\
4362.503 & $W$ & $16.8$ & 3\\
4362.896 & $W$ & $17.0$ & 4\\

\cutinhead{M33N 2009-01a}

4838.536 & $W$ & $17.0$ & 5\\
4839.475 & $W$ & $17.8$ & 5\\

\cutinhead{M33N 2010-10a}

5495.654 & $W$ & $18.1$ & 6\\

\enddata
\tablenotetext{a}{References: (1) \citet{qui06}; (2) \citet{ita06}; (3) \citet{nak07}; (4) \citet{kug07}; (5) \citet{nak09}; (6) \citet{yus10a}.}
\end{planotable}

\clearpage

\begin{planotable}{llrrrclr}
\tabletypesize{\scriptsize}
\tablenum{4}
\tablewidth{0pt}
\tablecolumns{8}
\tablecaption{M33 Novae\label{novatable}}
\tablehead{\colhead{} & \colhead{JD}&\colhead{$\Delta\alpha~cos\delta$\tablenotemark{a}} & \colhead{$\Delta\delta$\tablenotemark{a}} & \colhead{$a$\tablenotemark{a}}&\colhead{Discovery} & \colhead{} & \colhead{}  \\
\colhead{Nova} & \colhead{Discovery} & \colhead{($'$)} & \colhead{($'$)} & \colhead{($'$)} & \colhead{mag (Filter)} & \colhead{Type} & \colhead{References\tablenotemark{b}}}
\startdata
M33N 1919-12a    & 2422306.5 &   1.75 &   0.24 &   2.86 & 17.2 (pg) & \dots & 1 \\
M33N 1922-08a    & 2423292. &  -4.65 &   9.66 &  16.02 & 17.5 (pg) & \dots & 1 \\
M33N 1925-07a    & 2424348.5 &   5.04 &   9.44 &  10.79 & 17.9 (pg) & \dots & 1 \\
M33N 1925-12a    & 2424493.3 &   4.59 &  -5.26 &  11.62 & 18.1 (pg) & \dots & 1 \\
M33N 1927-07a    & 2425091.6 &  -5.73 &   7.26 &  15.15 & 17.7 (pg) & \dots & 1 \\
M33N 1927-09a    & 2425124.5 &   0.50 &  -1.96 &   2.71 & \dots     & \dots & 1 \\
M33N 1928-10a    & 2425534.5 &  -1.68 &  -8.45 &   9.00 & 16.0 (pg) & \dots & 1 \\
M33N 1949-08a    & 2433157.5 & -20.48 & -10.77 &  31.69 & 16.6 (pg) & \dots & 1 \\
M33N 1955-07a    & 2435289.5 &  -3.54 &  -5.18 &   6.54 & 17.2 (V)  & \dots & 1 \\
M33N 1960-11a    & 2437253.51 &   3.46 &   0.84 &   5.52 & 16.4 (pg) & \dots & 1 \\
M33N 1961-03a    & 2437365.32 &  15.38 &  -3.77 &  28.03 & 18.5 (pg) & \dots & 1 \\
M33N 1961-11a    & 2437632.29 &   3.08 &  -6.06 &  10.27 & 18.0 (pg) & \dots & 1 \\
M33N 1962-09a    & 2437917.56 & -10.13 &   4.81 &  20.01 & 18.0 (pg) & \dots & 1 \\
M33N 1962-09b    & 2437929.53 &   0.94 &   1.94 &   2.16 & 17.3 (pg) & \dots & 1 \\
M33N 1969-11a    & 2440529.78 &   6.98 &   7.33 &  11.41 & 18.0 (V)  & \dots & 1 \\
M33N 1970-09a    & 2440836. &   7.42 &  17.53 &  19.04 & 18.0 (pg) & \dots & 1 \\
M33N 1974-12a    & 2442386.5 &  11.28 &  -6.81 &  23.37 & 16.3 (pg) & \dots & 1 \\
M33N 1975-10a    & 2442691.5 &  -3.21 &  -3.02 &   5.12 & 18.8 (pg) & \dots & 1 \\
M33N 1977-12a    & 2443506.5 &  -8.69 & -14.01 &  16.90 & 17.9 (pg) & \dots & 1 \\
M33N 1982-09a    & 2445229.41 &   3.13 &   6.46 &   7.20 & 17.9 (B)  & \dots & 1 \\
M33N 1986-10a    & 2446703.36 &  -3.27 &  -4.56 &   5.90 & 18.5 (B)  & \dots & 1 \\
M33N 1995-08a    & 2449955.5 &  -0.82 &   0.82 &   1.96 & 16.0 (Ha) & \dots & 1 \\
M33N 1995-08b    & 2449957.5 &  -7.26 &  -6.43 &  11.49 & 16.6 (Ha) & \dots & 1 \\
M33N 1995-09a    & 2450012.5 &   3.44 &  -2.32 &   7.32 & 16.2 (Ha) & \dots & 1 \\
M33N 1996-12a    & 2450426.5 &   6.41 &  -4.96 &  14.16 & 19.1 (Ha) & \dots & 1 \\
M33N 1997-09a    & 2450692.5 &   7.20 &   3.10 &  11.21 & 19.3 (Ha) & \dots & 1 \\
M33N 2001-11a    & 2452229.26 &  -1.74 &   1.25 &   3.76 & 16.5 (w)  & \dots & 1 \\
M33N 2003-09a    & 2452883.9 &   0.06 &   0.22 &   0.23 & 16.9 (w)  & Fe IIb? & 2 \\
M33N 2006-09a    & 2454006.7 &  -6.92 &  10.21 &  19.67 & 16.6 (w)  & Fe II    & 3 \\
M33N 2007-09a    & 2454362.13 &   1.66 &  17.96 &  19.78 & 16.2 (R)  & He/N     & 3,4 \\
M33N 2008-02a    & 2454523.97 &   5.49 &  -0.14 &   9.34 & 16.5 (w)  & Fe II    & 5 \\
M33N 2009-01a    & 2454839.04 &  -2.26 & -13.91 &  14.95 & 17.0 (w)  & He/N      & 3 \\
M33N 2010-07a    & 2455376.5 &  -3.02 &   6.75 &  10.90 & 17.1 (w)  & \dots     & 1 \\
M33N 2010-10a    & 2455496.15 &   1.33 &   6.27 &   6.66 & 17.7 (w)  & Fe IIb    & 3 \\
M33N 2010-11a    & 2455528.03 &  -5.04 & -22.14 &  23.47 & 16.1 (w)  & Fe IIb   & 3 \\
M33N 2010-12a    & 2455547.92 &   5.82 &   4.36 &   9.06 & 16.4 (w)  & He/N      & 3 \\
\enddata
\tablenotetext{a}{Offsets from the nucleus ($a$ is the semimajor axis of the elliptical isophote passing through the position of the nova).}
\tablenotetext{b}{References: (1) positions and magnitudes from \citet{pie10}; (2) \citet{sch03}; (3) this work; (4) \citet{wag07}; (5) \citet{dim08}.}
\end{planotable}

\clearpage

\begin{deluxetable}{lcccccc}
\tabletypesize{\scriptsize}
\tablenum{5}
\tablewidth{0pt}
\tablecolumns{6}
\tablecaption{Balmer Emission Line Properties\label{balmerline}}
\tablehead{
\colhead{} & \multicolumn{2}{c}{EW (\AA)} & \multicolumn{2}{c}{FWHM (km~s$^{-1}$)\tablenotemark{a}} & \colhead{} & \colhead{} \\
\colhead{Nova} & \colhead{H$\beta$} & \colhead{H$\alpha$} & \colhead{H$\beta$} & \colhead{H$\alpha$} & \colhead{Type} & \colhead{References\tablenotemark{b}}}
\startdata
M33N 2003-09a  & \dots  & \dots & \dots & 2700:\tablenotemark{c} & Fe IIb? & 1 \\
M33N 2006-09a  & $-85$  & $-190$ & 1420 & 1370 & Fe II & 2 \\
M33N 2007-09a  & $-160$ & $-760$& 4260 & 4800 & He/N & 2 \\
M33N 2008-02a  & \dots  & \dots & \dots & 1700\tablenotemark{d} & Fe II & 3 \\
M33N 2009-01a (narrow)  & \dots  & \dots  & 1670 & 2510 & He/N & 2 \\
M33N 2009-01a (broad)   & \dots  & \dots  & 7220 & 5970 & He/N & 2 \\
M33N 2010-10a  & $-270$ & $-1630$ & 5060 & 4210 & Fe IIb & 2 \\
M33N 2010-11a  & $-110$ & $-220$ & 2800 & 2610 & Fe IIb & 2 \\
M33N 2010-12a  & $-170$ & $-1020$ & 3860 & 4070 & He/N & 2 \\
\enddata
\tablenotetext{a}{Estimated uncertainty $\pm 100$~km~s$^{-1}$.}
\tablenotetext{b}{References: (1) \citet{sch03}; (2) this work; (3) \citet{dim08}.}
\tablenotetext{c}{FWHM estimate based on reported 5400 km~s$^{-1}$ FWZI of H$\alpha$.}
\tablenotetext{d}{Reported FWHM of Balmer emission lines.}
\end{deluxetable}

\begin{planotable}{lcccr}
\tabletypesize{\scriptsize}
\tablenum{6}
\tablewidth{0pt}
\tablecolumns{5}
\tablecaption{Light Curve Parameters\label{lcparam}}
\tablehead{\colhead{Nova} & \colhead{Filter} & \colhead{$M_{max}$} & \colhead{Fade Rate (mag d$^{-1}$)} & \colhead{$t_2$ (days)}}
\startdata
M33N 2006-09a &$B$ & $ -8.12\pm0.15$& $0.059\pm0.003$ & $  33.7\pm1.7$ \\
              &$V$ & $ -8.46\pm0.15$& $0.059\pm0.003$ & $  33.8\pm1.8$ \\
M33N 2007-09a &$B$ & $ -8.53\pm0.15$& $0.183\pm0.006$ & $  11.0\pm0.3$ \\
              &$V$ & $ -8.93\pm0.15$& $0.342\pm0.013$ & $   5.9\pm0.2$ \\
M33N 2009-01a &$B$ & $ -7.65\pm0.15$& $0.120\pm0.007$ & $  16.7\pm1.0$ \\
              &$V$ & $ -7.99\pm0.15$& $0.168\pm0.008$ & $  11.9\pm0.5$ \\
M33N 2010-10a &$B$ & $ -6.55\pm0.15$& $0.086\pm0.005$ & $  23.1\pm1.3$ \\
              &$V$ & $ -6.89\pm0.15$& $0.098\pm0.006$ & $  20.4\pm1.2$ \\
\enddata
\end{planotable}

\end{document}